\documentclass{article}

\usepackage{frascatiphys}
\usepackage{graphicx}
\usepackage{amsmath}
\usepackage{amssymb}
\usepackage{tabularx}
\usepackage{booktabs}

\begin{document}

\title{
NUCLEAR PARTON DISTRIBUTION FUNCTIONS
}
\author{
Petja Paakkinen        \\
{\em University of Jyvaskyla, Department of Physics, P.O. Box 35, FI-40014 University of Jyvaskyla, Finland}
}
\maketitle
\baselineskip=10pt
\begin{abstract}
I review recent developments in the extraction of nuclear parton distribution functions. First describing the global analysis framework, I then present a comparison of the latest analyses in terms of included data and theoretical details, pointing out a few general trends.
\end{abstract}

\baselineskip=14pt

\section{Introduction: Collinear factorization and nuclear PDFs}
Parton distribution functions (PDFs) are defined in the context of collinear factorization theorem, which states that when a hard scale $Q^2$ is involved, the hard-process cross section for the colliding hadrons $A$, $B$ to produce a final state $k$ (in association with anything else) can be factorized in terms of a sum over the involved partons $i$, $j$ as
\begin{equation}
  {\rm d}\sigma^{AB \rightarrow k+X} \; \overset{\vspace{-1cm} Q \gg \Lambda_{\rm QCD}}{=} \; \sum_{i,j,X'} \, f_i^A(Q^2) \otimes {\rm d}\hat{\sigma}^{ij \rightarrow k+X'}(Q^2) \otimes f_j^B(Q^2) \; + \; {\cal O}(1/Q^2)
\end{equation}
up to power corrections in the reciprocal of the hard scale\cite{Collins:1989gx}. By virtue of the asymptotic freedom of QCD, the coefficient functions ${\rm d}\hat{\sigma}^{ij \rightarrow k+X'}$ are perturbatively calculable but the PDFs $f_i^A$, $f_j^B$ contain long-range physics and cannot be obtained by perturbative means. However, the PDFs are universal, process independent, and obey the DGLAP equations
\begin{equation}
  Q^2 \frac{\partial f_i}{\partial Q^2} = \sum_j P_{ij} \otimes f_j
\end{equation}
with splitting functions $P_{ij}$ governing the scale evolution\cite{DGLAP}.

For a nucleus $A$ with $Z$ protons and $N = A - Z$ neutrons, it is convenient to write
\begin{equation}
  f_i^A(x,Q^2) = \frac{Z}{A} f_i^{{\rm p}/A}(x,Q^2) + \frac{N}{A} f_i^{{\rm n}/A}(x,Q^2),
\label{eq:fullnuc}
\end{equation}
where $f_i^{{\rm p}/A}$ are the PDFs of a bound proton and the neutron contents $f_i^{{\rm n}/A}$ are obtained from $f_i^{{\rm p}/A}$ via isospin symmetry. As revealed by deep inelastic scattering (DIS) experiments, the bound nucleon PDFs are not the same as those of a free proton, but are modified in a nontrivial way. This observation has lead to global analyses of nuclear parton distribution functions (nPDFs); for earlier reviews, see Refs.\cite{Eskola:2012rg,Paukkunen:2014nqa,Paukkunen:2017bbm}. The first such fit was EKS98\cite{Eskola:1998iy} also including Drell--Yan (DY) dilepton data, followed by HKM\cite{Hirai:2001np} providing the first error analysis. Both of these were done in leading order (LO) perturbative QCD; the first next-to-leading order (NLO) analysis was provided by nDS\cite{deFlorian:2003qf}. The EPS08 analysis\cite{Eskola:2008ca} was the first to include RHIC dAu hadron-production data.

\begin{figure}[tb]
    \begin{center}
        {\includegraphics[width=87mm]{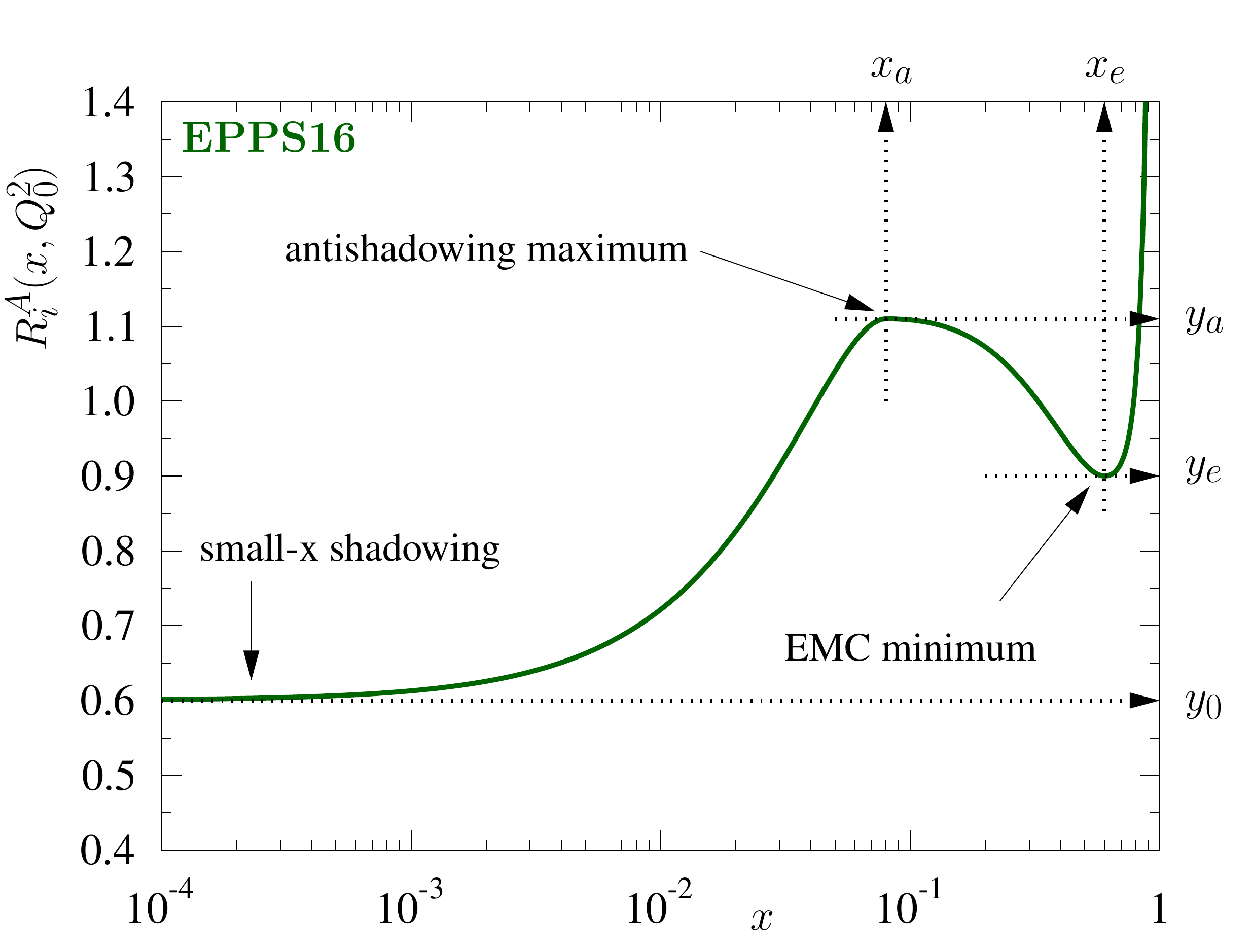}}
        {\includegraphics[width=87mm]{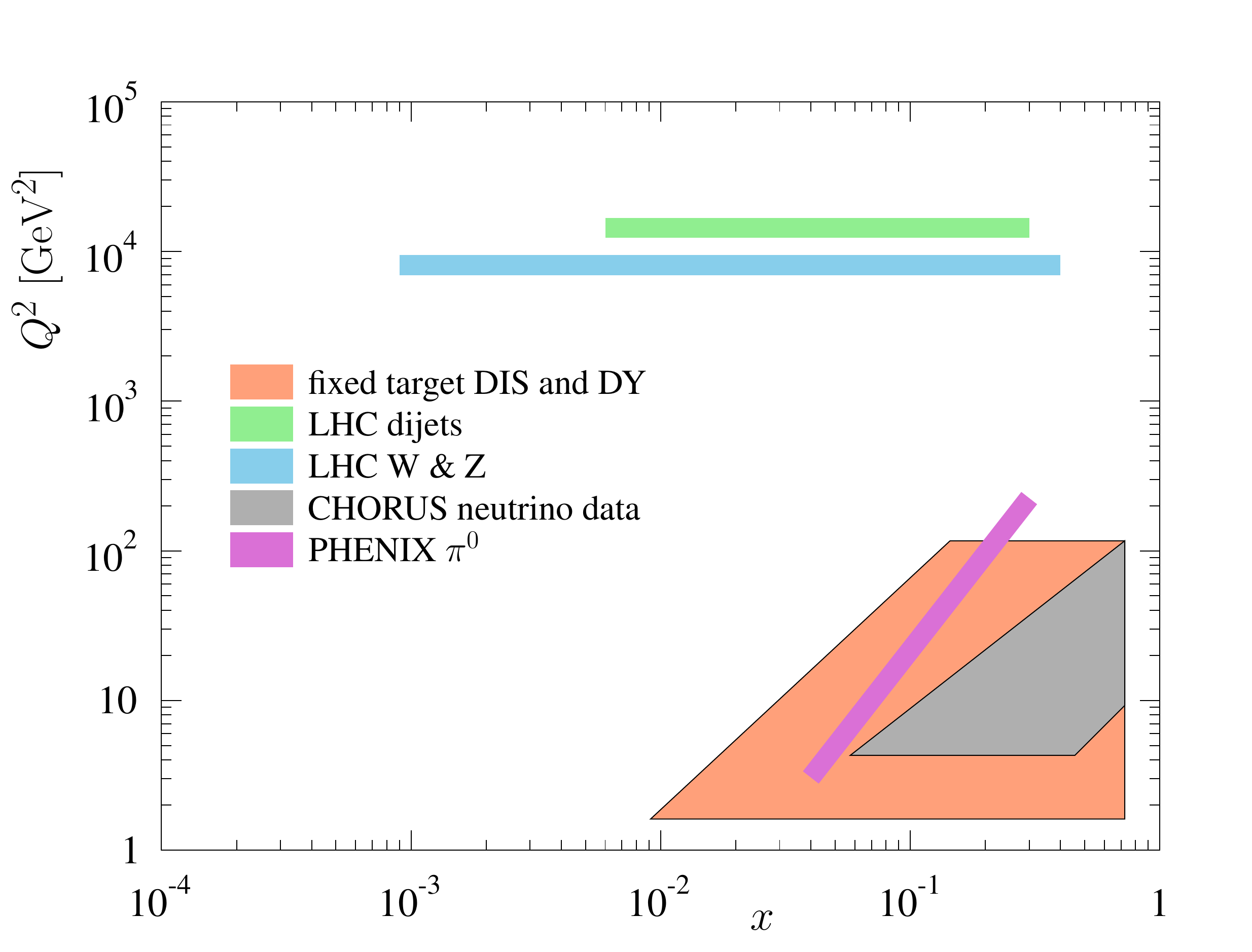}}
        \caption{\it Left: Typical form of PDF modifications in a nucleus. Right: Kinematic reach of the data used in nPDF global analyses. Figures from Ref.\cite{Eskola:2016oht}.}
        \label{fig:xQ2}
    \end{center}
\end{figure}

The relation of the bound-proton PDFs with respect to free-proton PDFs $f_i^{\rm p}$ is often expressed in terms of the nuclear modification factors
\begin{equation}
  R_i^A(x,Q^2) = \frac{f_i^{{\rm p}/A}(x,Q^2)}{f_i^{\rm p}(x,Q^2)}.
\label{eq:nucmod}
\end{equation}
A typical form of such modifications is shown in the left panel of Fig.~\ref{fig:xQ2}: small-$x$ shadowing followed by antishadowing, EMC-effect, and Fermi motion at large $x$. The global-analysis procedure is the same as in free-proton fits (see Ref.\cite{Gao:2017yyd} for a review), but there is a further complication since not enough data are available to fit each nucleus independently, and instead one needs to parametrize also the mass number dependence. Also the kinematic reach of the available data is more restricted than in corresponding free proton fits; see the right panel of Fig.~\ref{fig:xQ2} for an illustration of the data used in the most recent EPPS16 analysis\cite{Eskola:2016oht}.

\section{Global analysis}
\begin{figure}[tb]
    \begin{center}
        {\includegraphics[height=89mm]{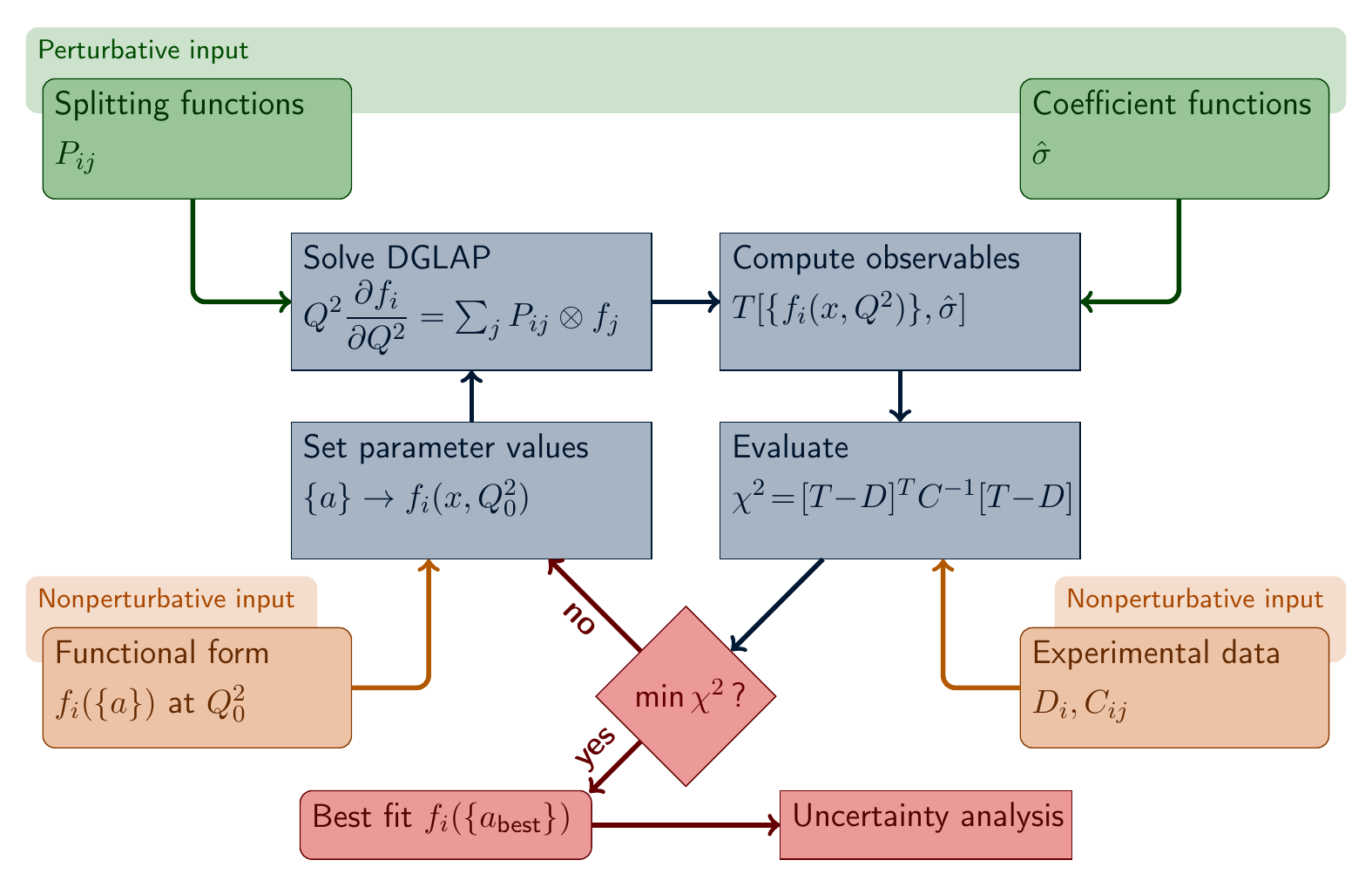}}
        \caption{\it Flowchart for PDF global analysis.}
        \label{fig:ga}
    \end{center}
\end{figure}
The PDF global analysis aims at finding the best possible parameter values such that a large set of experimental data from various hard processes are optimally described. In practice this is done by minimizing the figure-of-merit function
\begin{equation}
  \chi^2_\text{global} = \sum_{i,j}\,[T_i(\{a\}) - D_i]\,C^{-1}_{ij}\,[T_j(\{a\}) - D_j] \label{eq:chi2}
\end{equation}
with respect to a set of parameters $\{a\}$. Fig.~\ref{fig:ga} summarizes the various steps and inputs needed in the minimization process. One begins by parameterizing the PDFs at some initial scale $Q_0^2$ and sets initial values for the parameters. The PDFs are then evolved to higher scales by solving the DGLAP equations and then convoluted with the coefficient functions to obtain theory predictions $T_i$. These are then compared to experimental values $D_i$ with covariances $C_{ij}$ using Eq.~\eqref{eq:chi2}. If the minimum attainable $\chi^2$ was reached, one declares that the best fit was found and proceeds to uncertainty analysis. If not, one alters the parameter values and computes the observables again. Since this loop has to be traversed multiple times, fast methods for both solving the DGLAP equations and computing the observables are needed so that neither of these becomes a bottleneck in the analysis.

It is worth to note that all of the inputs in Fig.~\ref{fig:ga} are possible sources of uncertainty. However, the theoretical uncertainties related to the choice of the parametrization form or neglecting higher order corrections in the splitting and coefficient functions are hard to quantify. Thus one usually restricts oneself to asking how the experimental uncertainties translate to uncertainties in the parameter values. The Hessian method for uncertainty extraction\cite{Pumplin:2001ct} relies on the quadratic approximation of the $\chi^2$ function
\begin{equation}
  \chi^2_\text{global} \approx \chi^2_0 + \sum_{i,j}\,(a_i - a_i^0)\,H_{ij}\,(a_j - a_j^0) = \chi^2_0 + \sum_{i} z_i^2, \label{eq:chi2quad}
\end{equation}
where $\chi^2_0$ is the minimum of $\chi^2$ at parameter values $a_i^0$, and the latter expression is written in terms of new parameters $z_i$ such that they are linear combinations of the original parameters and uncorrelated in the quadratic approximation. In these new parameters, one then finds the maximal upward and downward deviations $\delta z^\pm_i$ corresponding to a fixed increase $\Delta\chi^2$ in the $\chi^2_{\rm global}$ function.

To enable a general user to calculate PDF related uncertainties, global analyses provide ``error sets'', PDFs evaluated with the parameter deviations $\delta z^\pm_i$ corresponding to the tolerance $\Delta\chi^2$. The uncertainty of any PDF related quantity $X$ can then be obtained separately for the upward and downward directions with
\begin{equation}
  \left( \delta X^\pm \right)^2 = \sum_i \left[ \substack{\max \\ \min} \left\{ X\left(\delta z^+_i\right)-X_0, X\left(\delta z^-_i\right)-X_0, 0 \right\} \right]^2,
\end{equation}
where $X_0$ is the value obtained using PDFs with best fit parameters, the ``central set'', and $X\left(\delta z^\pm_i\right)$ are calculated with the error sets. The allowed error tolerance $\Delta\chi^2$ varies from analysis to analysis, as do the details of how to extract this value. A common practice is to use a ``90\% confidence criterion'', where $\Delta\chi^2$ is taken to be the average of changes in $\chi^2$ corresponding to a maximal shifts in each of the new parameters $z_i$ such that all data sets remain within their 90\% confidence ranges. For a more detailed discussion, see Ref.\cite{Eskola:2016oht}.

\section{Nuclear PDF comparison}

\begin{table}[t]
\centering
\caption{ \it Selection of global nPDF analyses. Table adapted from Ref.\cite{Paukkunen:2017bbm}.}
\vskip 0.1 in
\newcolumntype{Y}{>{\centering\arraybackslash}X}
    \begin{tabularx}{175mm}{|c|Y|Y|Y|Y|Y|}
      \hline
      & EPS09\cite{Eskola:2009uj} & DSSZ\cite{deFlorian:2011fp} & KA15\cite{Khanpour:2016pph} & nCTEQ15\cite{Kovarik:2015cma} & EPPS16\cite{Eskola:2016oht} \\
      \hline
      \hline
      Order in $\alpha_s$ & LO \& NLO & NLO & NNLO & NLO & NLO \\
      NC DIS $l$A/$l$d & \checkmark & \checkmark & \checkmark & \checkmark & \checkmark \\
      DY pA/pd & \checkmark & \checkmark & \checkmark & \checkmark & \checkmark \\
      RHIC pions dAu/pp & \checkmark & \checkmark & & \checkmark & \checkmark \\
      $\nu$A DIS & & \checkmark & & & \checkmark \\
      $\pi$A DY & & & & & \checkmark \\
      LHC pPb W, Z & & & & & \checkmark \\
      LHC pPb jets & & & & & \checkmark \\
      & & & & & \\
      $Q$ cut in DIS & 1.3 GeV & 1 GeV & 1 GeV & 2 GeV & 1.3 GeV \\
      datapoints & 929 & 1579 & 1479 & 708 & 1811 \\
      {free parameters} & 15 & 25 & 16 & 16 & 20 \\
      error analysis & Hessian & Hessian & Hessian & Hessian & Hessian \\
      error tolerance $\Delta\chi^2$ & 50 & 30 & not given & 35 & 52 \\
      Free proton PDFs & CTEQ6.1 & MSTW2008 & JR09 & CTEQ6M-like & CT14 \\
      HQ treatment & ZM-VFNS & GM-VFNS & ZM-VFNS & GM-VFNS & GM-VFNS \\
      Flavour separation & no & no & no & valence & full \\
      Weight data in $\chi^2$ & yes & no & no & no & no \\
      \hline
    \end{tabularx}
\label{tab:npdfs}
\end{table}
Now that we are familiar with the global analysis framework, it is time to compare results of different analyses. Table~\ref{tab:npdfs} summarizes the details of the latest global nPDF analyses, including EPS09\cite{Eskola:2009uj}, DSSZ\cite{deFlorian:2011fp}, KA15\cite{Khanpour:2016pph}, nCTEQ15\cite{Kovarik:2015cma} and EPPS16\cite{Eskola:2016oht}. Most of these are NLO QCD analyses. While the KA15 analysis was performed at next-to-NLO (NNLO), they only included DIS and DY data, thus lacking a direct constraint for gluons, and are not at the same global footing as other (NLO) analyses which also include inclusive pion production data from RHIC. DSSZ were the first to include neutrino--nucleus DIS, but the full potential of these data was not fully unleashed due to an assumption of flavour symmetric valence and sea quark nuclear modifications. Independent valence distributions were first allowed in nCTEQ15, but with very limited constraints since no $\nu$A data were included. Most recently, EPPS16 provided the first analysis with parametric freedom for all flavours and constraints not only from $\nu$A DIS, but also $\pi$A DY and LHC pPb W and Z production. Due to lack of sufficient statistics, the latter observables however are not able to give as stringent constraints as $\nu$A DIS. Also new in EPPS16, more constraints for gluon nuclear modifications were obtained from the inclusion of LHC pPb dijet data. This has enabled EPPS16 to lift the data weight which was used in the EPS09 analysis to emphasize the impact of RHIC pion data in the absence of other gluon constraints. An important development is the employment of the general-mass variable-flavor-number scheme (GM-VFNS), see Ref.\cite{Thorne:2008xf} and the references therein, for heavy-quark treatment in DSSZ, nCTEQ15 and EPPS16 as opposed to the zero-mass scheme (ZM-VFNS) used in EPS09 and KA15.

\begin{figure}[tb]
  \begin{center}
    \includegraphics[width=175mm]{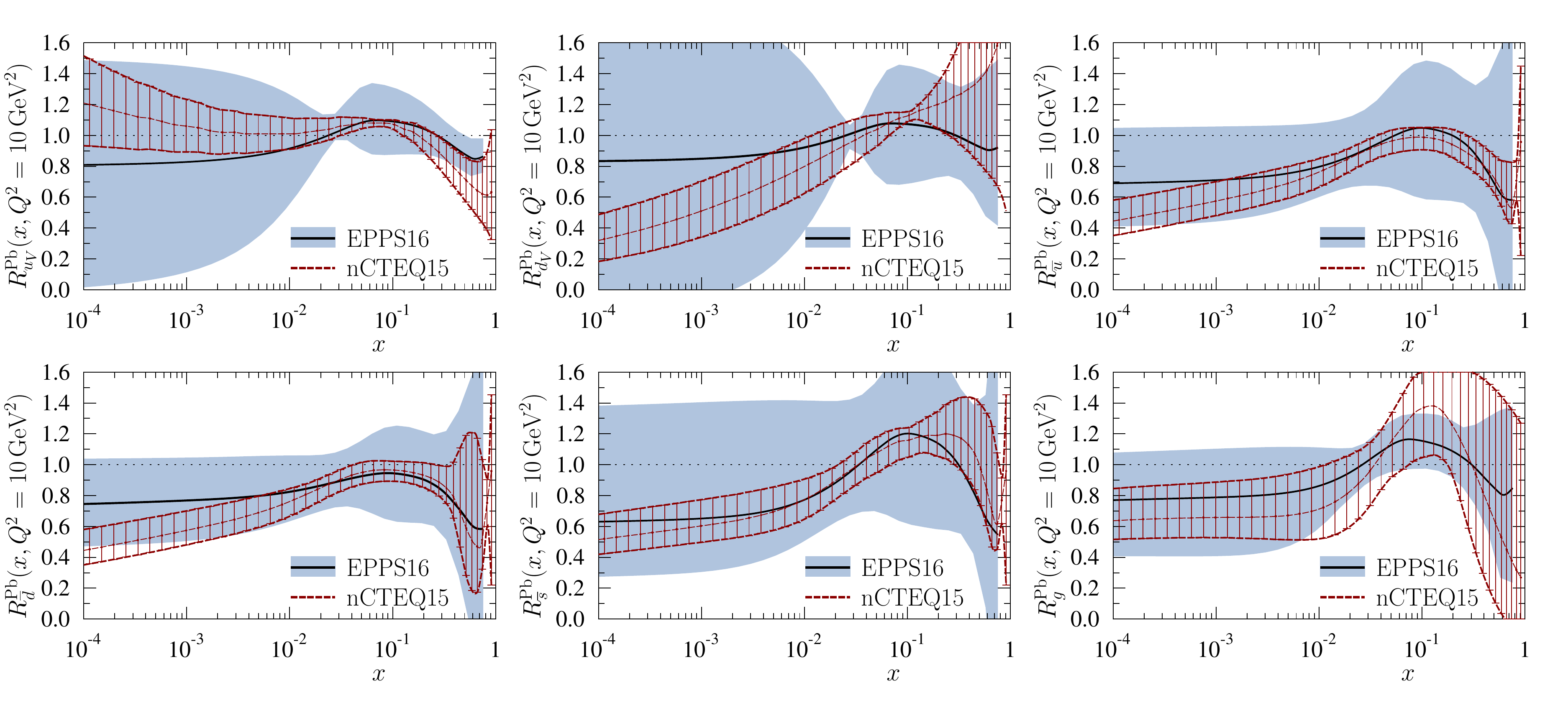}
    \caption{\it Comparison of the EPPS16 and nCTEQ15 nuclear PDFs. Figure from Ref.\cite{Eskola:2016oht}.}
    \label{fig:nCTEQ}
  \end{center}
\end{figure}
\begin{figure}[tb]
  \begin{center}
    \includegraphics[width=60mm]{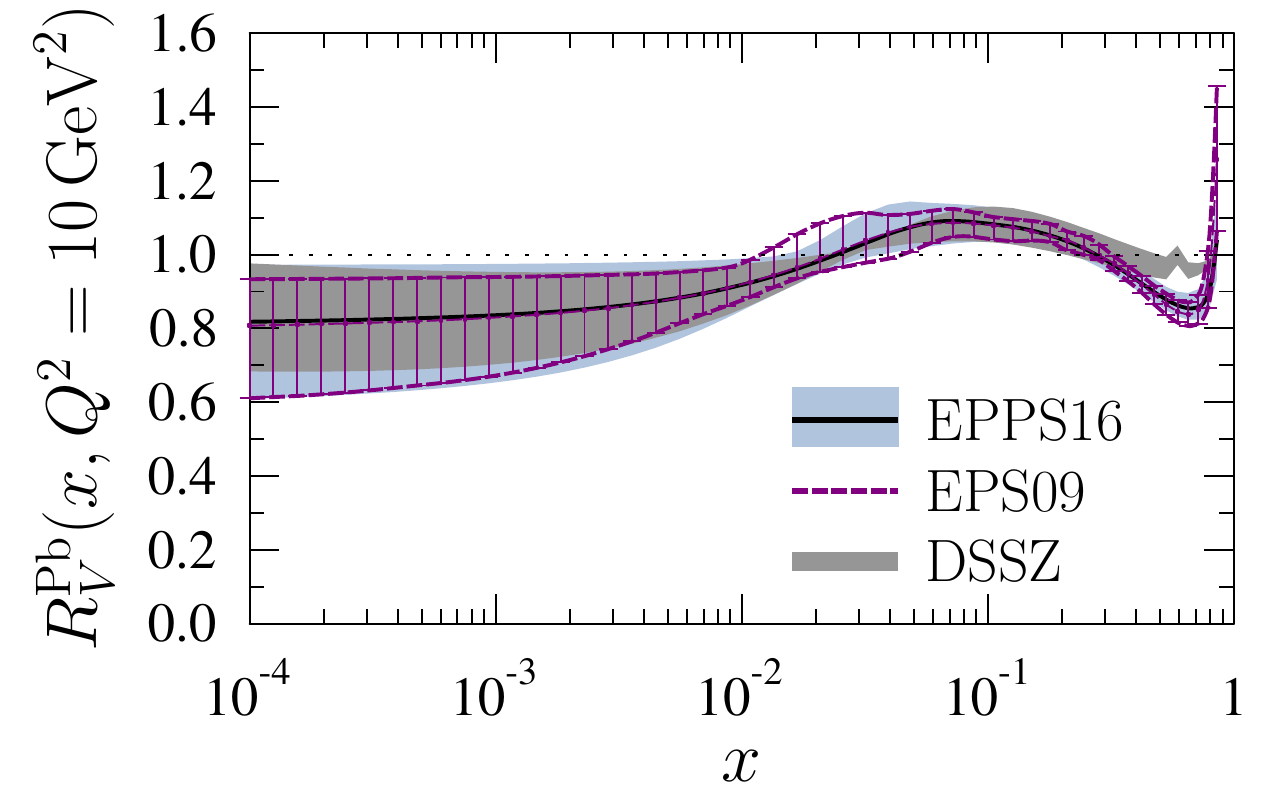}
    \hspace{-3mm}
    \includegraphics[width=60mm]{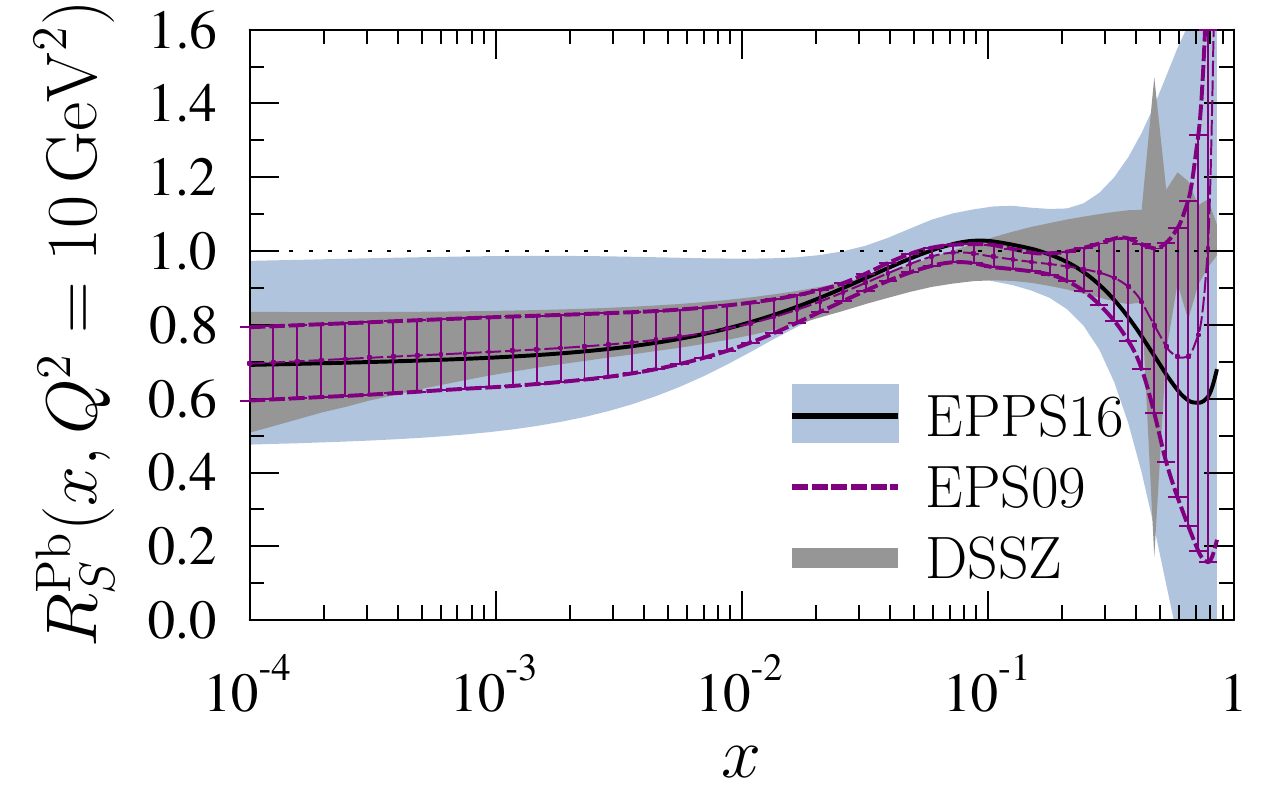}
    \hspace{-3mm}
    \includegraphics[width=60mm]{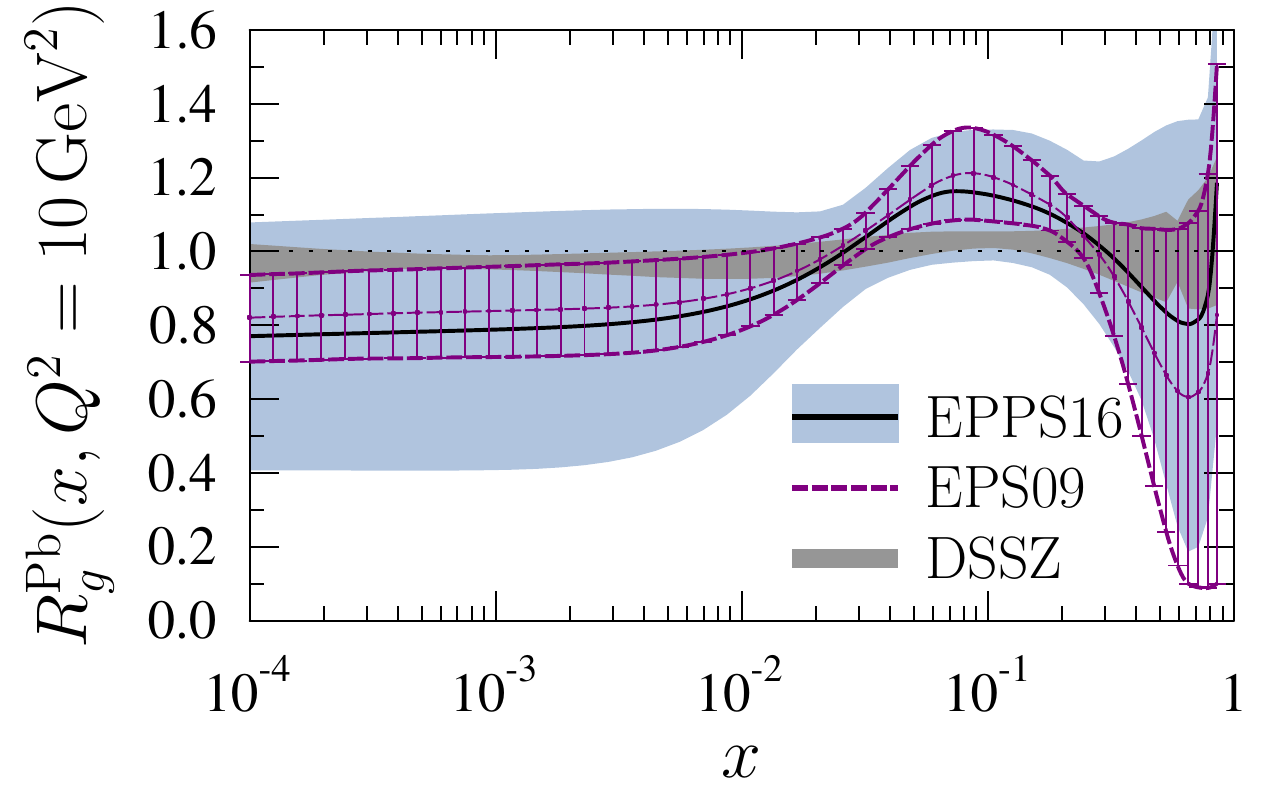}
    \caption{\it Comparison of the EPPS16, EPS09 and DSSZ nuclear PDFs. Figure from Ref.\cite{Eskola:2016oht}.}
    \label{fig:DSSZ}
  \end{center}
\end{figure}

Fig.~\ref{fig:nCTEQ} shows the nuclear modifications of partons in lead nucleus from EPPS16 and nCTEQ15 analyses. The two are compatible as the error bands always overlap, but there are certain differences which need to be addressed. First, the central predictions for valence-quark modifications obtained by the two analyses appear quite different. While in EPPS16 the $u$ and $d$ valence quark modifications are very similar, in nCTEQ15 these differ significantly with $u$ quark exhibiting a large EMC suppression whereas $d$ quark obtains an enhancement in the same kinematic region. This is possibly due to nCTEQ15 using isospin-symmetric DIS data and having no $\nu$A DIS in their fit. As pointed out also in Ref.\cite{Kovarik:2015cma}, such differences become more dilute when we construct the PDFs of the full nucleus according to Eq.~\eqref{eq:fullnuc}. This is also the reason why EPPS16 valence uncertainties are so large: while the average valence quark distribution is well under control (cf.~Fig.~\ref{fig:DSSZ} left panel), we would need high-precision data on non-isoscalar nuclei to constrain the difference in $u$ and $d$ modifications. Second, the EPPS16 sea-quark uncertainties are much larger than those of nCTEQ15. This is simply due to nCTEQ15 having less freedom in their parametrization: in nCTEQ15 there are only 2 free parameters for all sea quarks with no flavour dependence, whereas EPPS16 has altogether 9 free sea-quark parameters, of which only 3 are common to all flavours. Third, the nCTEQ15 gluon uncertainties at high $x$ are larger than those of EPPS16 resulting from nCTEQ15 having a harder $Q^2$ cut in DIS data and not including LHC jet data.

Comparing EPPS16 with EPS09 and DSSZ in Fig.~\ref{fig:DSSZ}, since the latter have no flavour freedom, we find it sensible only to compare the averages
\begin{equation}
    R_{\rm V}^{\rm Pb} \equiv  \frac{u^{\rm p/Pb}_{\rm V}+d^{\rm p/Pb}_{\rm V}}{u^{\rm p}_{\rm V}+d^{\rm p}_{\rm V}},
    \qquad
    R_{\rm S}^{\rm Pb} \equiv  \frac{\overline{u}^{\rm p/Pb}+\overline{d}^{\rm p/Pb}+\overline{s}^{\rm p/Pb}}{\overline{u}^{\rm p}+\overline{d}^{\rm p}+\overline{s}^{\rm p}}.
\end{equation}
The valence-quark modifications of these three analyses are very similar to each other, except in the EMC region, where DSSZ is close to unity. This has been identified with a misinterpretation of the isospin corrections in the DSSZ analysis\cite{Paukkunen:2014nqa}. The EPPS16 sea-quark uncertainty is larger than in EPS09 and DSSZ due to additional parametric freedom from allowing flavour separation, but the shape of the obtained modifications match very well. Regarding the gluon modifications, we find EPS09 and EPPS16 to give similar results. The EPS09 uncertainties however are artificially small because the additional weight for RHIC data was used. DSSZ gives a rather different behavior, as it contains virtually no gluon modifications at all. This follows from the choice in DSSZ to use nuclear fragmentation functions (nFFs), the gluonic component of which was constrained with the very same pion production data as used in the DSSZ analysis. Hence, by necessity, they arrived with similar small gluon modifications as in nDS which was used in the nFF extraction\cite{Eskola:2012rg}.

\section{Conclusions}

I have reviewed the recent nuclear-PDF analyses and the developments therein. A major step forward is the inclusion of LHC pPb data. Especially the gluon-PDF extraction is benefiting from the new constraints coming from the dijet measurements. For electroweak pPb data to give stringent constraints, we need to wait until measurements with better statistics are published. Apart from including more and more data, we can see a few general trends which can be expected to continue also in the future. Most prominently, we are experiencing a shift towards parameterizing the full flavour dependence of nPDFs, as opposed to using simplifying assumptions. While this tends to make flavour by flavour uncertainties larger at first, it renders the global analysis more data driven and thus less biased. Also, the treatment of heavy-quark mass effects with GM-VFNS is becoming a well established practice. An emerging development seems to be the inclusion of NNLO corrections; the pace at which these will be implemented in the future analyses remains to be seen.

\section{Acknowledgements}
I thank K.~J.~Eskola and H.~Paukkunen for comments. Financial support from the Magnus Ehrnrooth Foundation and the Academy of Finland, Project 297058, is acknowledged.

\end{document}